# ORIGAMIX, a CdTe-based spectro-imager development for nuclear applications

S. Dubos [a], H. Lemaire [b], S. Schanne [a], O. Limousin [a], F. Carrel [b], V. Schoepff [b], C. Blondel [a]

[a] CEA, DSM/IRFU, Astrophysics Division, 91191 GIF-SUR-YVETTE Cedex, France

[b] CEA, LIST, Sensors and Electronic Architectures Laboratory, 91191 GIF-SUR-YVETTE Cedex, France

## Abstract

The Astrophysics Division of CEA Saclay has a long history in the development of CdTe based pixelated detection planes for X and gamma-ray astronomy, with time-resolved imaging and spectrometric capabilities. The last generation, named Caliste HD, is an all-in-one modular instrument that fulfils requirements for space applications. Its full-custom front-end electronics is designed to work over a large energy range from 2 keV to 1 MeV with excellent spectroscopic performances, in particular between 10 and 100 keV (0.56 keV FWHM and 0.67 keV FWHM at 13.9 and 59.5 keV).

In the frame of the ORIGAMIX project, a consortium based on research laboratories and industrials has been settled in order to develop a new generation of gamma camera. The aim is to develop a system based on the Caliste architecture for post-accidental interventions or homeland security, but integrating new properties (advanced spectrometry, hybrid working mode) and suitable for industry. A first prototype was designed and tested to acquire feedback for further developments.

The main difference between astronomy and nuclear applications is the radiation flux and energy-ranges involved. In this study, we particularly focused on spectrometric performances with high energies and high fluxes. Therefore, our device was exposed to energies up to 700 keV ($^{133}$Ba, $^{137}$Cs) and we measured the evolution of energy resolution (0.96 keV at 80 keV, 2.18 keV at 356 keV, 3.33 keV at 662 keV). Detection efficiency decreases after 150 keV, as Compton effect becomes dominant. However, CALISTE is also designed to handle multiple events, enabling Compton scattering reconstruction, which can drastically improve detection efficiencies and dynamic range for higher energies up to 1408 keV ($^{22}$Na, $^{60}$Co, $^{152}$Eu) within a 1 mm-thick detector. In particular, such spectrometric performances obtained with $^{152}$Eu and $^{60}$Co were never measured before with this kind of detector.

## 1. Introduction

Severe accidents are part of the recent history of the nuclear industry. The two most famous are the Chernobyl disaster in 1986 and the Fukushima event in 2011. These accidents had a huge impact on human health or environment, and still have substantial societal consequences. In most cases, accidental events and post-accidental events require some human interventions in order to mitigate the crisis. For this reason, adapted and performing tools are necessary to minimize the radiological impact on operators.

Localization of potentially strongly irradiating radioactive hot spots is a major issue in a crisis situation. Their position and nature are unknown because of the accidental configuration of the environment. Gamma cameras are well-adapted tools for this topic and enable hot spot localization by superimposition of gamma and visible images. Industrial gamma imaging systems have been available for decades (for instance, CARTOGAM [1], RADCAM [2]) and some recent developments have been brought to the market (for instance, GAMPIX [3, 4, 5], POLARIS-H [6]). Moreover, there is an active research on this topic in the scientific community [7].

The purpose of this article is to present recent advances dedicated to the development of a high performance coded aperture imaging spectrometer. The latter will be able to locate and to discriminate radionuclides close in energy. This system will be based on the Caliste HD technology [8] developed by the Astrophysics Division of CEA DSM (a division of the French Atomic Energy Commission).

This work is carried out within the framework of the ORIGAMIX Project, which is funded by a Grant from the French Government, coordinated by the French National Research Agency as a part of the "Investissements d'Avenir" Program, under reference ANR-11-RSNR-0016.

In a first part, we will present the Caliste HD imaging spectrometer, used as a baseline for the ORIGAMIX project. A first demonstrator of a portable system will be introduced. In a second part, we will use this prototype to evaluate its spectrometric performances in a large energy range up to 1.4 MeV, by taking advantage of different operating modes implemented in Caliste. After the conclusion, we will expose some perspectives regarding imaging capabilities and source recognition.

## 2. ORIGAMIX project

### 2.1. Baseline: Caliste technology

The Astrophysics division of CEA DSM has been involved for more than ten years in a continuous development effort to create high-performance detection areas for hard X-ray and gamma-ray astronomy. Designed to be installed at the focal plane of a space telescope, these instruments have to provide both time-resolved imaging and spectrometric capabilities in the 2-250 keV energy range ([9], [10]). Cadmium-telluride-based semiconductor compounds are a good choice for radiation detectors operating in these wavelengths, as they combine wide band gap, high density and high atomic number, suitable to emphasize photoelectric interaction. Moreover, this kind of material can be operated at room temperature or with moderate cooling, with remarkable spectrometric performances [11].

The Caliste HD imaging spectrometer is the third version of this development series [8]. It uses a 1 mm-thick Schottky CdTe detector divided into 256 pixels disposed in a 16x16 matrix, with a pixel pitch of 625 μm. Each pixel of the 1 cm$^2$ detection surface is connected to an independent spectroscopic channel of a specifically-developed front-end electronics. This electrical body, placed below the detector part, includes 8 mixed analog-digital IDeF-X HD ASICs [12] and multiplexing interfaces. The output signal is extracted from the device via a 4x4 pin grid array located on the opposite side of the active surface (Fig. 1). With such a characteristic modular arrangement, a single Caliste HD device is aboutable on its four sides and can be used as a baseline to create large detection areas with a high pixel density and spectrometric capabilities [13]. The total power consumption of this device is also very low: only 200 mW (0.8 mW per channel).

Caliste HD has been qualified in a laboratory environment with a $^{241}$Am source, and shown excellent energy resolutions - 562 eV FWHM and 666 eV FWHM at respectively 13.9 and 59.5 keV (equivalent to 4.0 and 1.1% energy resolution) [13]. According to theoretical calculations conducted with a Fano factor of 0.15 for the CdTe sensor, minimum absolute resolutions are 226 eV FWHM at 13.9 keV and 467 eV at 59.5 keV. Experimental measurements of energy resolutions thereby correspond to an electronic noise level around 46 and 50 e⁻ rms: this performance is due to the very low noise of the electronics as well as low input capacitance and low leakage current from the detector part (values below 1 pF and 1 pA, respectively).

Moreover, a low-level detection threshold of 1.3 keV was recently reached with this device. This value gives the ability to perform imaging and spectrometry measurement of low energy photons starting from 2 keV, with a minimal detection efficiency as high as 40% [14].

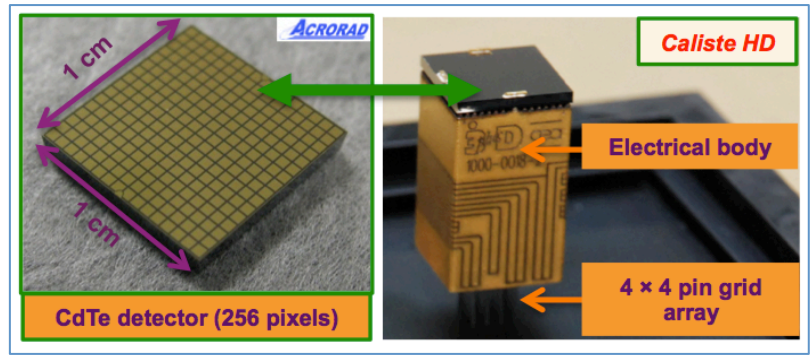

*Figure 1: Overview of a Caliste HD module, with its 1 mm-thick CdTe detector subdivided into 256 pixels, the electrical body and data links.*

As the Caliste HD micro-camera was initially developed to be a sufficiently mature candidate for future X or gamma-ray astronomy missions, it is also an appropriate choice for post-accident interventions such as those studied during the ORIGAMIX project. With its 256 self-triggered pixels, this module is suitable for precise source localization. These imaging capabilities can be optimized between available field of view and angular resolution using various coded masks placed in front of the entrance window. The key advantage of Caliste HD compared to other solutions is to add highly resolved spectrometric information to these images, potentially able to identify and discriminate different radionuclides present in the field of view of the system, especially with gamma emissions in close energy ranges.

As previous versions of Caliste cameras have an energy upper limit of 250 keV, the front-end electronics of Caliste HD can be set to 4 different tunable gains, one of them allowing an extended dynamic energy range from 2 keV to 1 MeV. Such energy range is consistent with nuclear application requirements, and will be used for the following experiments.

### 2.2 First demonstrator

Caliste HD was initially operated in laboratory conditions and required to be installed in a vacuum chamber with an active cooling system at approximately -10°C. Additional parts such as high-voltage filters, refrigerators, interface boards and data connections resulted in a large and heavy setup (example in Fig. 2 from [14]).

Within the framework of the ORIGAMIX project, we built a first demonstrator including all this material in a portable device (Fig. 2). The Caliste HD module is surrounded by a cooling system (Peltier coolers and cold plates), and placed behind a thin Beryllium window. Data links and power routing are added and the entire system is confined in a closed vessel. A static vacuum of typically $10^{-5}$ mbar is established with an external pump and can be maintained for hours. An external radiator is also added for thermal dissipation, and the detection module can be cooled down to approximately 0°C in this configuration.

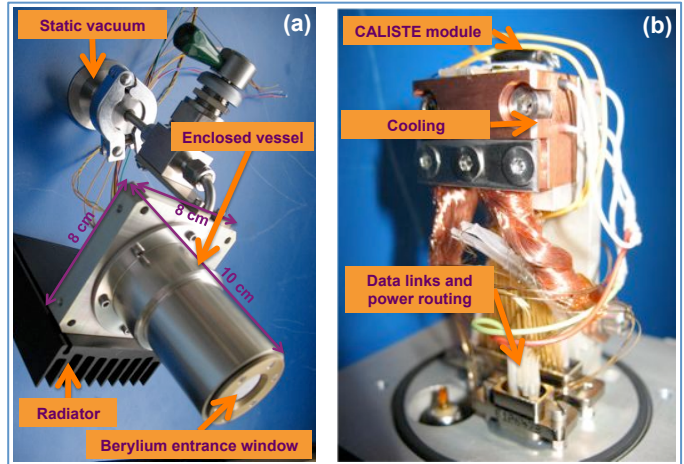

*Figure 2: (a) Overview of the first prototype developed in the frame of the ORIGAMIX project. This device is portable and only needs to be connected to a high-voltage source and a FPGA board for control and data extraction. (b) Detail of the assembly, with the Caliste HD unit, cooling systems and data / power links.*

### 2.3 Experimental setup

In order to evaluate spectrometric performances in a relevant environment, we placed our portable camera on a mobile rail with sources with activities between 11 and 74 MBq, from low to high energies such as $^{241}$Am - with its main peak at 59.5 keV - and $^{60}$Co which emits 1173.2 keV and 1332.5 keV gamma rays.

Different types of coded masks – widely used in the field of hard X-rays and gamma-rays detection as an indirect imaging technique [15] – were also tested during this measurement campaign in order to prepare for further imaging developments (Fig. 3).

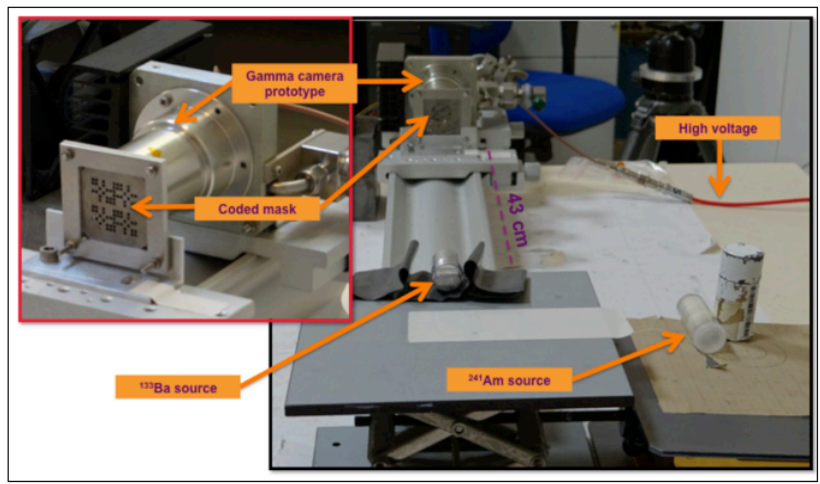

*Figure 3: Experimental setup configuration with the portable camera and a coded mask disposed on a mobile rail. Two different sources are placed in the field of view.*

The camera was connected to a FPGA board enabling a 14-bit ADC, used for slow-control and to extract signal amplitudes and time-tags corresponding to triggered pixels.

## 3. Spectrometric performances

### 3.1. Energy calibration and linearity

To work in the largest possible energy range, we selected the maximum dynamic range permitted by the electronics. A typical factor of 4 is applied in terms of gain between the lowest and the largest dynamic range (respectively up to 250 keV and up to 1 MeV), with only a slight increase of the electronic noise.

Due to gain and offset variations, each pixel of the entire matrix should be calibrated independently. If we combine acquisitions successively done with several different sources, we can identify characteristic photopeaks between 30 and 662 keV. Their energies are given in ADU (*Analog/Digital Units*) which depends on the ADC conversion factor. After displaying corresponding amplitudes of each photopeak, a linear function is used to model the energy response of each channel. Parameters such as gain and offset values are extracted and stored in a calibration matrix, which is used to derive the following results.

For the entire matrix, we found an average gain value of 51.47 eV/ADU. It is approximately 4 times larger than that measured in [14] using the minimum dynamic range, as expected. The gain standard deviation is found to be only 0.39 eV/ADU.

We also calculated the maximal *Integral Non-Linearity* (INL) for this energy range with an average value equal to ±0.81% (0-peak). If we consider the whole matrix, individual values of the INL never exceed 3%. Thus, 70% and 87% of the pixels have INL values respectively contained between 1 and 2%. Once again, these values are compatible with those found at the ASIC level [12], which illustrates the very good linearity of Caliste HD from low to high energies up to 662 keV.

### 3.2. Energy response for single events

In this part, we consider only "single" events – *i.e.* only one triggered pixel, without charge-sharing. With this choice, we limit the contribution of electronic noise in the reconstructed energy, as only one channel is involved to measure the energy of an impinging photon.

Using the calibration map, we build a global spectrum by summing all individual spectra obtained for each pixel. In Fig. 4, a superposition of sum spectrums obtained for various radioactive sources is displayed. Due to the thickness of the detector, we observe a progressive decrease of the detection efficiency (below 50% after 143 keV), with the Compton effect becoming predominant. Major energy peaks are indicated and also listed in Table 1.

In the same Table, we reported energy resolutions measured for these energy peaks. FWHM values are given for the entire matrix, over 256 pixels.

| Source | Energy | FWHM *Single-events* | Charge sharing | Expected efficiency |
|---|---|---|---|---|
| $^{241}$Am | 59.5 keV | 0.80 keV / 1.3% | 31.04% | 97.87% |
| $^{133}$Ba | 81.0 keV | 0.96 keV / 1.1% | 30.91% | 82.10% |
| $^{57}$Co | 122.1 keV | 1.10 keV / 0.9% | 33.11% | 50.43% |
| $^{133}$Ba | 356.0 keV | 2.18 keV / 0.6% | 55.33% | 7.63% |
| $^{22}$Na | 511 keV | 3.40 keV / 0.7% | 61.00% | 5.13% |
| $^{137}$Cs | 661.7 keV | 3.33 keV / 0.5% | 61.57% | 4.35% |

*Table 1: Detail of energy resolutions measured over 256 pixels for single-events and for various energies from 60 to 662 keV (absolute and relative values). The charge-sharing ratio is measured in an energy window covering each photopeak. Expected efficiencies are calculated with linear attenuation coefficients and thicknesses of each material layer, considering all interactions (photoelectric and Compton) [14].*

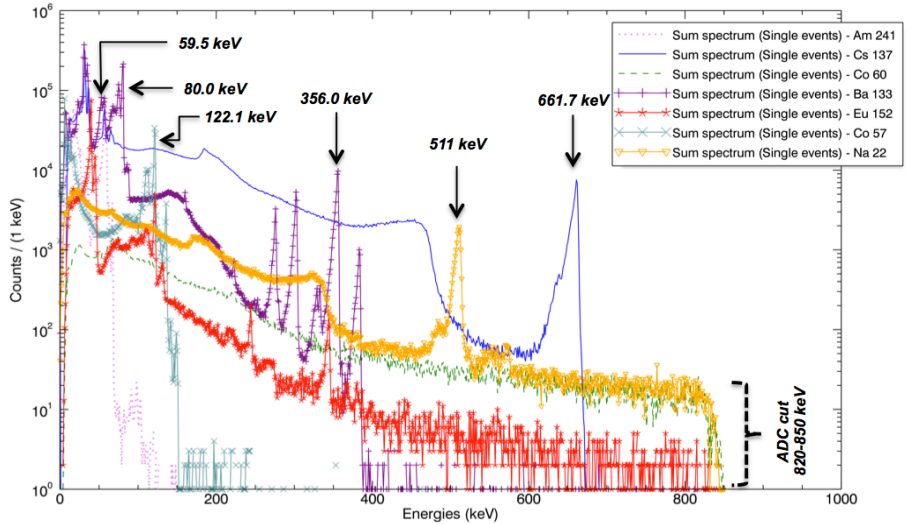

*Figure 4: Superimposition of sum spectrums obtained for various radioactive sources with photopeak energies up to 662 keV. Only single events are kept to ensure best energy resolutions. The energy cut at 820-850 keV is due to an ADC limitation, with a partial measurement of emission spectrum for sources like $^{152}$Eu, $^{60}$Co or $^{22}$Na.*

The 14-bit ADC used for this first test campaign limits maximum measurable energies to below 820-850 keV. As this limit does not come from the Caliste HD module itself, another ADC with a larger dynamic should help using the full potential of the IDeF-X HD ASICs and reach single events measured up to 1 MeV.

### 3.3. Charge sharing

The Caliste HD architecture gives the possibility to monitor the number of triggered pixels and fractions of deposited energies for each incoming photon. Energy deposition over multiple neighbouring pixels strongly depends on the detector geometry. A simple model proposed by Iniewski *et al.* [16] predicts a charge-sharing rate of ≈22%, and has already been confirmed with low energy measurements ([14], [17]). This model does not take into account energy deposition processes into the detector volume, and only considers e$^-$/h diffusion within the CdTe material for charge-sharing calculation.

As this limitation can be negligible for the lowest energies, we can see in Table 1 measured charge-sharing ratios for increasing energy bands. This value is doubled between 60 keV and 662 keV (respectively 31.04 and 61.57%), and can be explained by the increase of the charge cloud diameter created by a single photon interacting with the detector. The predominance of Compton effect for high-energy interactions is thus responsible of such high probability to share energy over multiple pixels.

By summing energies deposited in each hit pixels, we can reconstitute the energy of the original photon. This has consequences on energy resolutions, as electronic noise of multiple spectroscopic channels is quadratically summed. At the same time, a slight energy loss due to an incomplete charge collection in the interpixel gap is responsible for a tail on the left of the photopeak and a small shift of the peak toward low energies (Fig. 5 and Fig. 6).

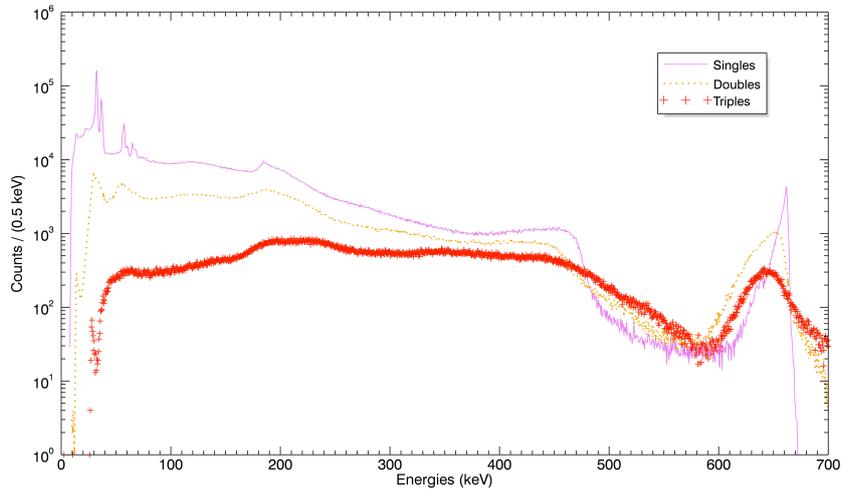

*Figure 5: Sum spectrums obtained with a $^{137}$Cs source by selecting events caused by respectively 1, 2 or 3 triggered neighbouring pixels. Energy resolutions are equal to 3.33 keV for single events, 15.69 keV (double events) and 39.38 keV (triple events). Energy losses in the photopeak for double and triple events are respectively 1.31% and 2.67%.*

### 3.4. Extended dynamic range up to 1.4 MeV

While the reconstruction of shared events degrades the global energy resolution, it is also a good method to increase the detection efficiency by taking all triggers into account, and especially Compton effect. If we apply such a reconstruction with our previously measured sources, we can extend the available energy window to energies as high as 1.4 MeV (Fig. 6). As no single-events are measurable above 850 keV, the charge-sharing rate is equal to 100%.

Energy resolutions obtained for all multiplicities with the entire pixel matrix are listed in Table 2, for the highest energies considered. These values are of course worse than those obtained in Table 1, but still sufficient to ensure future source discrimination capabilities.

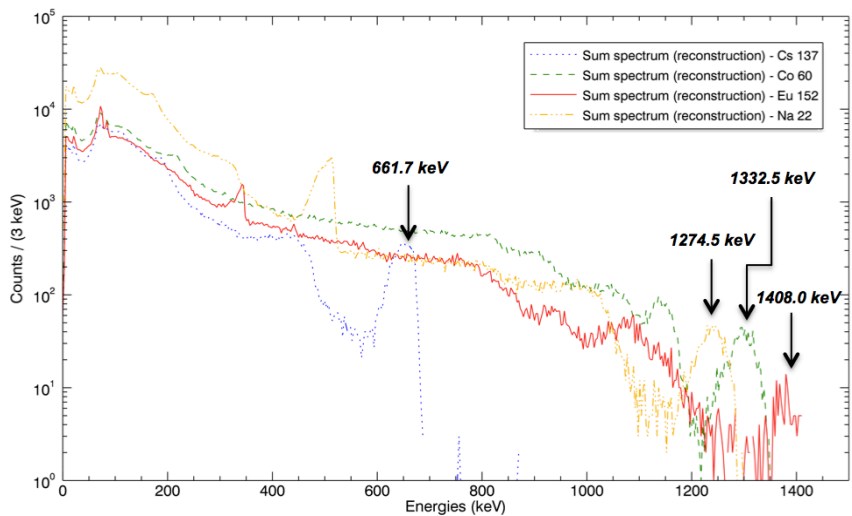

*Figure 6: Superposition of sum spectrums obtained for various radioactive sources with photopeak energies up to 1.4 MeV. We consider all multiplicities to rebuild charge-sharing and Compton events, which also allows overcoming the ADC or ASIC high limit for single events measurement.*

| Source | Energy | FWHM – *All multiplicities* | Expected efficiency |
|---|---|---|---|
| $^{137}$Cs | 661.7 keV | 13.94 keV / 2.11% | 4.35% |
| $^{22}$Na | 1274.5 keV | 58.09 keV / 4.56% | 2.88% |
| $^{60}$Co | 1332.5 keV | 59.95 keV / 4.50% | 2.82% |
| $^{152}$Eu | 1408.0 keV | 52.08 keV / 3.70% | 2.74% |

*Table 2: Detail of energy resolutions measured over 256 pixels for reconstructed multiple events, for various energies from 662 to 1408 keV with total expected efficiencies calculated within a 1 mm-thick CdTe detector.*

## 4. Conclusion and perspectives

In these first runs, we demonstrated that Caliste HD could be integrated in a portable and modular device, without loosing its excellent spectrometric properties, even with large energy ranges. In particular, we can take advantage of its operating mode to balance its different properties. We can thus ensure best energy resolutions with single events, at the price of a reduced energy window and lower efficiencies. By considering all multiplicities and reconstructing Compton interactions, we can maximize the efficiency and the available dynamic range up to 1.4 MeV. Even with limited detection efficiencies due to the 1 mm-thick CdTe crystal used, Caliste HD can take advantage of high fluxes generally found in nuclear physics, in comparison with astrophysics. These properties constitute major advantages for developments planned within the framework of the ORIGAMIX project.

By taking advantage of the spectroscopic performances of the Caliste HD module, the imaging properties of this device could be used to localize multiple sources of different energies within its field of view. To demonstrate imaging capabilities, we performed a first test with a radioactive source, placed 1 m away from the detector, which can be considered as point-like in this setup. The source used was $^{241}$Am with 74 MBq activity. The detector was equipped with a coded mask of 32×32 pixels (of 1 mm each), with an open fraction of 40%, manufactured out of a 300 μm-thick Ta-foil. This coded-mask imaging system offers a very large total field of view, covering ±64° × ±64°, which can be represented by a matrix of 68×68 pixels (the size of the detector pixels viewed from the mask-detector distance). The central pixels in the field of view have an angular size of 3.6°. Taking into account the mask and detector pixel sizes, the system has a theoretical angular resolution of 6.7°, due to the small mask-detector distance (1 cm), which on the other hand allows for the very large field of view.

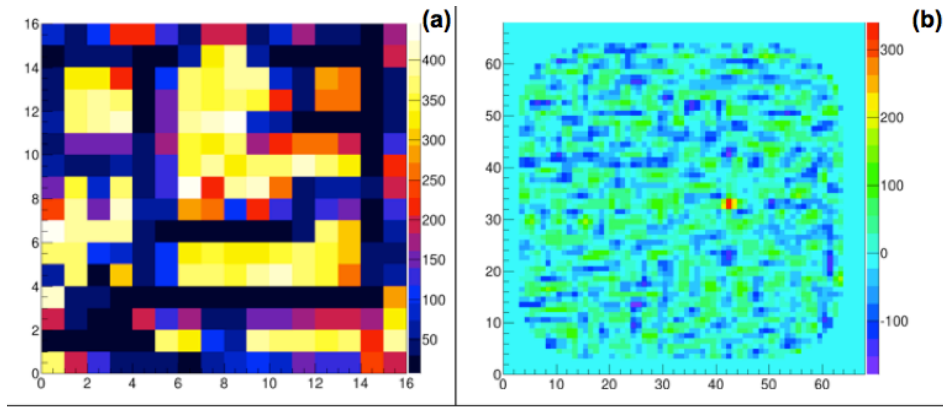

*Figure 7: (a) Detector plane images in the 60 keV energy band obtained with a $^{241}$Am source used in the setup. (b) Reconstructed field of view image showing the signal to noise ratio (number of sigma) of the reconstructed source position. At variance with astrophysical cases, our setup employs high count-rate sources in a low background environment. This image shows therefore a high level of coding noise, which however does not harm the source localization.*

In an exposure of 3850 seconds, the detector registered 72500 single counts in the 55 to 63 keV energy band covering the $^{241}$Am photopeak (Fig. 7a). The $^{241}$Am source location was reconstructed using the imaging software developed for ECLAIRs and LOFT [18,19], which is based on the deconvolution of the recorded detector image by the coded mask pattern.

The source was reconstructed with a high confidence level, at pixel (X,Y)=(42,32.5), corresponding to the actual setup (Fig. 7b). The discussion of these algorithms and their applications for this development go beyond the scope of this article and will be postponed for a future publication. Nevertheless, we can conclude that on such a small portable camera, the deconvolution method, developed in our lab for space applications, seems appropriate to localize point-like sources in its field of view.

Other image reconstruction methods, based on expectation-maximization algorithms, are currently under study by our group and will be also presented in a forthcoming article, as well as a detailed discussion about source discrimination capabilities.